\pdfoutput=1

\documentclass[11pt,twoside,a4paper,cmspaper,final,collab]{cms-tdr}
\def\svnVersion{222095}\def\cmsCernNo{2013-169}\def\cmsCernDate{\today}\def\cmsMessage{Published in Physical Review Letters as \href{http://dx.doi.org/10.1103/PhysRevD.88.112009}{\doi{10.1103/PhysRevD.88.112009.}}}

\begin{document}\cmsNoteHeader{SMP-12-004}

\hyphenation{had-ron-i-za-tion}
\hyphenation{cal-or-i-me-ter}
\hyphenation{de-vices}

\RCS$Revision: 216812 $
\RCS$HeadURL: svn+ssh://svn.cern.ch/reps/tdr2/papers/SMP-12-004/trunk/SMP-12-004.tex $
\RCS$Id: SMP-12-004.tex 216812 2013-11-17 23:20:06Z alverson $
\newcommand{\Vvar}{\ensuremath{V}\xspace}
\newcommand{\costhe}{\ensuremath{\cos\theta^{*}}\xspace}
\newcommand{\ydif}{\ensuremath{y_\text{dif}}\xspace}
\newcommand{\ysum}{\ensuremath{y_\text{sum}}\xspace}
\newcommand{\yj}{\ensuremath{y_\text{jet}}\xspace}
\newcommand{\ypho}{\ensuremath{y_{\gamma}}\xspace}
\newcommand{\yz}{\ensuremath{y_\cPZ}\xspace}
\newcommand{\yv}{\ensuremath{y_\Vvar}\xspace}
\newcommand{\pz}{\ensuremath{p_{\mathrm{z}}}\xspace}
\newcommand{\mur}{\ensuremath{\mu_{\mathrm{R}}}\xspace}
\newcommand{\muf}{\ensuremath{\mu_{\mathrm{F}}}\xspace}
\newcommand{\ptgam}{\ensuremath{\pt^{\gamma}}\xspace}
\newlength\cmsFigWidth
\ifthenelse{\boolean{cms@external}}{\setlength\cmsFigWidth{0.4\textwidth}}{\setlength\cmsFigWidth{0.4\textwidth}}
\ifthenelse{\boolean{cms@external}}{\providecommand{\cmsLeft}{top}}{\providecommand{\cmsLeft}{left}}
\ifthenelse{\boolean{cms@external}}{\providecommand{\cmsRight}{bottom}}{\providecommand{\cmsRight}{right}}

\cmsNoteHeader{SMP-12-004} 

\title{\texorpdfstring{Rapidity distributions in exclusive \cPZ\ + jet and $\cPgg$~+~jet events in $\Pp\Pp$ collisions at $\sqrt{s} = 7\TeV$}{Rapidity distributions in exclusive Z + jet and photon + jet events in pp collisions at sqrt(s)=7 TeV}}

\date{\today}

\abstract{Rapidity distributions are presented for events containing either a
Z boson or a photon with a single jet in proton-proton
collisions produced at the CERN LHC. The data, collected with the CMS
detector at $\sqrt{s}=7\TeV$, correspond to an integrated luminosity of 5.0\fbinv. The individual rapidity distributions of the boson and the jet are
consistent within 5\% with expectations from perturbative QCD. However,
QCD predictions for the sum and the difference in rapidities of the two
final-state objects show discrepancies with CMS data. In
particular, next-to-leading-order QCD calculations, and two common Monte Carlo
event generators using different methods to match matrix-element
partons with parton showers, appear inconsistent with the data
as well as with each other.}

\hypersetup{
pdfauthor={CMS Collaboration},
pdftitle= {Rapidity distributions in exclusive Z + jet and photon + jet events in pp collisions at sqrt(s)=7 TeV},
pdfsubject={CMS},
pdfkeywords={CMS, physics, Z boson, photon, angular distribution, QCD, perturbative test, matrix element, parton shower matching, MCFM, SHERPA, MADGRAPH}}

\maketitle

From the time of Rutherford's first scattering experiments, measuring angular
distributions has provided a tool for understanding the structure and
interactions of matter.
Measurements of the rapidity distributions in \Vvar  + jet events,
where \Vvar  refers either to a \cPZ\ boson or a photon, can provide an
important check of quantum chromodynamics (QCD) and event generators
used to simulate elementary processes.
For \cPZ\ boson decays into \EE or \MM, triggering is very efficient and nearly
background-free,  and
from a theoretical point of view, the presence of the electroweak vertex makes
the perturbative calculation of dynamical quantities even more robust.
Since next-to-leading-order (NLO) perturbative QCD calculations are available for \cPZ\ bosons produced in association with four or fewer jets~\cite{z4jets}, as well as
for $\gamma$ + jet production~\cite{Owens:1982,Owens:1987, jetphox},
detailed comparison with data is possible.
In addition, a precise understanding of these processes is also required in
searches for new physics and in studies of the Higgs boson, for which \cPZ\ + jets events
constitute an important background.

The rapidity of a particle is defined as
$y=(1/ 2) \ln[(E + \pz)/(E - \pz)]$, where $E$ is the energy and
\pz is the momentum component along the direction of the counterclockwise
circulating proton beam.
The invariant rapidity difference can be written in terms of the measured
quantities \yv and \yj as  $\ydif= \abs{\yv - \yj}/2$.
The quantity $\ysum=\abs{\yv + \yj}/2$ is the boost from the laboratory frame to
the center-of-mass frame of the \Vvar  and jet.
In the laboratory frame, \yv and \yj are highly correlated because \Vvar  + jet
production usually involves a relatively high-momentum valence quark
interacting with a low-momentum  gluon or antiquark, which results in events
where the \Vvar  and jet are usually on the same end of the detector.
The rapidities  \ysum and  \ydif  are effectively rotations in phase space of the \yv
and  \yj system that yield two approximately uncorrelated quantities.
The distribution in \ysum  depends mainly on the parton distribution functions (PDF),
while the distribution in \ydif  reflects the leading order (LO) partonic differential cross section.
Distributions in the \ysum and \ydif quantities were measured previously at
$\sqrt{s} = 1.96\TeV$ by the D0 Collaboration~\cite{Dzero:2010a}.
Related angular quantities in \Vvar  + jet events have been measured at the LHC by
CMS~\cite{CMS-EventShape} and
ATLAS~\cite{ATLAS-ZJet,ATLAS-NewZJet,ATLAS-WJet,ATLAS-PhotonAngular}.
In this Letter, we compare theoretical predictions for normalized distributions in $\abs{\yv}$,
$\abs{\yj}$, \ysum, and \ydif with data collected by the Compact Muon Solenoid
(CMS) experiment.

The kinematic properties of \cPZ\ + jets events at the
Tevatron~\cite{Dzero:2010a} were found to be well-described by the NLO \MCFM
program~\cite{mcfm:2002,mcfm:2003}. For \cPZ\ + jets production at the LHC, \MCFM provides
predictions together with estimates of their uncertainties.
For $\gamma$ + jet events, the program developed by Owens~\cite{Owens:1987} is
used for NLO predictions and their uncertainties.
This 
calculation employs
fragmentation functions~\cite{jetphox} to parametrize small-angle photon
emission. Previous studies of the transverse momentum (\pt) distributions of
photons showed agreement between LHC data and a variety of QCD
predictions~\cite{CMS:2011dp,Chatrchyan:2011ue,ATLAS-InclusivePho,ATLAS-InclusivePhoSecond}.
In all \MCFM calculations, both the renormalization (\mur)
and factorization (\muf) scales are set to the invariant mass of the lepton
pair. For NLO prompt photon calculations, the scales are set to the \pt of the
photon.

Programs that use matrix-element descriptions of jet systems are accurate
only when the partons have large transverse momentum or are well separated,
while event generators using parton showers perform well in describing soft and
small-angle radiation~\cite{Mrenna:2003if}.
Hybrid methods are used to combine matrix-element prescriptions and
parton showers to optimize performance for all regions of phase space.
These programs generally proceed in three stages:
(i) calculating the lowest-order ``tree-level" contribution,
(ii) simulating parton showering and clustering of final-state partons into jets,
and (iii) employing one of two schemes to minimize double-counting of matrix-element jets and those produced
by parton showering.
The MLM~\cite{MLM:2006} procedure rejects events when showering changes the event topology, while the CKKW
method~\cite{CKKW:2001} uses a weighting scheme based on shower history.
A previous comparison of hybrid methods~\cite{Alwall:2008} found large
differences in distributions of \ydif between the MLM and CKKW methods for W + jet
production. Hybrid models have been compared with \Vvar  + jet data at the
Tevatron~\cite{Dzero:2010a}, where the CKKW method implemented in \SHERPA
provided the best overall description of the observed distributions in \ysum
and \ydif, but with a significantly different cross section. We also
compare predictions from the two hybrid event generators with our
\Vvar + jet data using \MADGRAPH 5.1.1.0~\cite{MADGRAPH}, which implements the
MLM scheme, and \SHERPA 1.3.1~\cite{SHERPA}, which uses the CKKW method.
For \MADGRAPH, matching scales are chosen to be  20\GeV for \cPZ\ bosons and 9--12\GeV for photons.
The \mur and \muf scales are set to
$\sqrt{  {m_{\cPZ}^2 + \Sigma_\mathrm{jets} \pt^2 }}$ and
$\sqrt{ { \left(\ptgam\right)^2+ \Sigma_\mathrm{jets} \pt^2 }}$
for \cPZ\ bosons and photons, respectively, where $m_{\cPZ}$ is the mass of
the \cPZ\ boson and \ptgam is the photon transverse momentum.
The \PYTHIA 6.4.24 event generator is used for parton showers and
hadronization~\cite{Sjostrand:2006kx}.
For the \SHERPA events, the matching scales are 20 and 10\GeV for \cPZ\ bosons and
photons, respectively. The $\muf$  and $\mur$ scales are set to
$m_{\cPZ}$ and \ptgam for \cPZ\ bosons and photons, respectively.
The parton-shower module \textsc{apacic++}~2.0~\cite{Apacic} is used before
the \PYTHIA hadronization procedure.
In our comparison, the \SHERPA simulations use the NLO CTEQ6.6M~\cite{CTEQ66M} PDF.
The use of different order PDF in hybrid calculations is disputed~\cite{NLOLOPDF}: both LO and NLO PDF have
been used in theory~\cite{Alwall:2008} and experiments~\cite{ ATLAS-WJet, Dzero:2010a}.
To investigate this effect, the \MADGRAPH
simulation is studied using both the LO and NLO CTEQ6~\cite{CTEQ6} PDF.

The central feature of the CMS apparatus is a superconducting solenoid of
6\unit{m} internal diameter, providing a magnetic field of 3.8\unit{T}. A
silicon pixel and strip tracker, a lead tungstate crystal electromagnetic
calorimeter, and a brass/scintillator hadron calorimeter reside within the
magnetic field volume. Muons are detected in gas-ionization detectors embedded
in the steel of the flux-return yoke of the magnet. In addition to the barrel
and endcap detectors, CMS has extensive forward calorimetry. A more detailed
description of CMS is given in Ref.~\cite{CMS:detector}. A right-handed
coordinate system is defined in CMS, with the origin at the center of the
detector, the $x$ axis pointing to the center of the LHC ring, and the $y$ axis
perpendicular to the plane of the LHC.
The polar angle $\theta$ is measured from the positive $z$ axis and the
azimuthal angle $\phi$ is measured in the $x$-$y$ plane in radians.
Pseudorapidity, which is given by $\eta = - \ln[\tan(\theta/2)]$, is used for specifying
acceptance requirements.

The data were collected during 2011 at a $\Pp\Pp$ center-of-mass energy
of 7\TeV, corresponding to an integrated luminosity of
$5.0\pm0.1\fbinv$~\cite{CMSLumi}.
Because of limitations in data handling, the triggers used for photon
candidates had to be partially suppressed, and the effective luminosity for
prompt-photon production was $4.9\pm0.1\pbinv$.
The multilevel trigger requires two electron or two muon
candidates, with respective minimum-\pt thresholds of 17 and 8\GeV, or 18 and
8\GeV, for the lepton of highest and next-highest \pt.  A photon candidate
is required to have $\ptgam > 30\GeV$.

Event reconstruction requires at least one vertex with $\abs{z}<15\cm$
located within the beam pipe (radius ${<}2\cm$).
Jets and leptons are reconstructed using the particle flow
algorithm~\cite{CMS:2010kk}, which classifies all stable particles
in an event using the full ensemble and redundancy of the CMS detector.
Jets are clustered using the anti-\kt  algorithm~\cite{antikt},
with a distance parameter of 0.5,  and are required to have
$\abs{\eta} < 2.4$ to assure good tracker coverage.
Jets,  which have typical energy-scale uncertainties of
${<}3\%$ and resolution better than $10\%$~\cite{CMS:2011jes}, are required to have $\pt > 30\GeV$.
The difference between actual and simulated resolution is ${<}1\%$, and simulations showed the
difference has a negligible effect on the rapidity distributions.
The energy of particles arising from additional overlapping $\Pp\Pp$
interactions in the same bunch crossing, but not associated with the hard
scattering, is referred to as ``pileup".
Pileup from charged particles is subtracted based on tracking
information from the other reconstructed vertices.
Neutral particle pileup contributes ${\approx}0.5\GeV$ to any jet for each
additional $\Pp\Pp$ interaction, and is subtracted from the jet energy.
The probabilities of observing pileup from additional hard interactions or
from double parton scattering are both ${<}1\%$ for \Vvar  + jet
events~\cite{Dzero:MPI,ATLASDPS}.
Jets below threshold are ignored, and if any other jet exceeds threshold,
the event is rejected.

Reconstructed \cPZ\ boson events are required to have at least two oppositely
charged leptons of the same flavor (electrons or muons), each with
$\pt > 20\GeV$ and $\abs{\eta} < 2.1$.
The pair is required to have an invariant mass in the range of 76--106\GeV
(close to $m_{\cPZ}$), and $\pt > 40\GeV$.
To include final-state radiation, the lepton energy is corrected
by adding all photon energy deposited within an ($\eta,\phi$) cone
of $\Delta R =\sqrt{(\Delta\eta)^2+(\Delta\phi)^2} =0.1$, centered on the track
direction.
The relative isolation of a lepton is defined by the sum of the scalar \pt of
all reconstructed particles (excluding the lepton) within a
($\eta,\phi$) cone of $\Delta R =0.4$ of the lepton direction,
divided by the lepton \pt.
Electron and muon candidates are required to have an isolation of less
than $20\%$ and $15\%$, respectively.
Leptons and jets are required to be separated by $\Delta R > 0.5$.
Detailed discussions of electron and muon reconstruction at CMS can be found
in Refs.~\cite{CMS:elepas,CMS:ele,CMS:muo}. Following these selections, the \cPZ\ + jet
sample is 99\% pure. The remaining contributions from \ttbar pairs, diboson
(\cPZ\cPZ, \PW\PW, or \PW\cPZ), and multijet processes are negligible~\cite{CMS_paper_WZ}.

Reconstructed photon candidates are required to have $\ptgam > 40\GeV$
to assure a fully efficient trigger, and $\abs{\eta} <1.44$ to avoid systematic
effects associated with crossing calorimeter boundaries.
Photons are reconstructed and selected as described in Ref.~\cite{CMS:2011dp},
using an isolation cone of $\Delta R=0.4$ in ($\eta,\phi$) space.
The isolation variables for photons are defined
 with the sum of the charged-particle scalar \pt required to be less than
2\GeV, and the sums of the electromagnetic and hadron calorimeter contributions
to be less than 4.2 and 2.2\GeV, respectively.
The background resulting from fragmentation of jets into collimated neutral
mesons that mimic a photon is estimated using the ``matrix" method of
Ref.~\cite{CMS_paper_WZ}.
The transverse spatial distribution of the energy in a cluster is used as a
discriminant.
Templates for the spatial transverse distributions of photon showers are taken
from \PYTHIA events, and reconstructed through the full CMS detector simulation
via \GEANTfour~\cite{geant4}.
The  templates for background are obtained from data using events
passing all selection requirements,  but with less stringent charged-particle
isolation criteria (set between 2 and 5\GeV).
Following the selection requirements, the fraction of photons in the sample is
determined individually for each bin in \ydif. The photon fraction decreases
from $(61\pm2)\%$ at $\ydif=0$ to $(36\pm5)\%$ at
$\ydif=1.4$, where the uncertainties are statistical.
To correct for background in the \ysum, $\abs{\yj}$, and $\abs{\ypho}$
distributions, events are weighted by the photon fractions as a function of
\ydif, while using the independence on \ysum to reduce point-to-point fluctuations.
The resulting effective fractions (photon purities) change by less than 12\% within the examined
ranges of \ysum, $\abs{\yj}$, and $|\ypho|$.

The reconstructed distributions are corrected for efficiency and resolution before determining the differential distributions in rapidity.
For \cPZ\ + jet events, efficiencies are evaluated from data and simulation
using a ``tag-and-probe" procedure introduced in Ref.~\cite{CMS_paper_WZ}.
The simulated spectra are scaled to match collision data as a function of the
lepton \pt and $\eta$, and are then used to compute the efficiency as a function
of the rapidity variables.
Photon efficiencies are obtained from simulation, and rescaled using the
measured electron efficiency, which is assumed to have the same $\eta$
dependence as photons. All rapidity distributions are corrected for the effects of
detector resolution using simulated events in an iterative unfolding
method~\cite{DAgostini:1994zf},
as implemented in the \textsc{RooUnfold} package~\cite{RooUnfold}.
For all rapidity variables, the size of the correction is smaller than $1\%$.
Only the distribution of $\abs{\yj}$ has significant bin migration due to effects
of resolution.
The other variables have a correction factor consistent with unity.

\begin{table}[ht]
\topcaption{
Summary of the relative experimental uncertainties in the yields for the bins
of largest $\abs{y}$. At smaller $\abs{y}$ values, these uncertainties are in general
smaller. The first three rows reflect uncertainties common to all three
analyses.
The following nine rows quantify the uncertainties by particle type according
to (i)~uncorrelated statistical uncertainty, (ii)~trigger and selection efficiency,
and (iii)~correlated estimation of the background,
separately for \EE, \MM, and $\gamma$ events.
}
\centering
\begin{scotch}{ lrrrr  }
Source	          		&$\abs{\yv}$ (\%)     		&$\abs{\yj}$ (\%)    		&\ysum (\%) 		&\ydif (\%)     	\\
\hline																		
Jet energy 		       	&0.4		&0.6		&0.3		&0.4	\\
Pileup reweight 		&0.5		&0.1		&0.3		&0.1	\\
Unfolding				&---		&5.0		&---		&---	\\
\hline
\Pe\ Statistical			&7.1		&2.2		&8.0		&5.7	\\
\Pe\ Efficiency		       	&3.1		&0.9		&3.2		&2.8	\\
\Pe\ Background 			&0.2		&0.2		&0.2		&0.2	\\
\hline
$\mu$ Statistical		&6.3		&1.9		&6.1		&4.6	\\
$\mu$ Efficiency	        	&1.5		&0.4		&1.2		&1.2	\\
$\mu$ Background 		&0.2		&0.2		&0.2		&0.2	\\
\hline
$\cPgg$ Statistical		&6.6		&19		&8.6		&15	\\
$\cPgg$ Efficiency		&0.4		&0.6 		&0.4		&1.0	\\ 
$\cPgg$ Background  	&7.0		&2.0		&1.0		&11	\\
\end{scotch}

\label{table:Uncert}
\end{table}

The sources and relative experimental uncertainties in the rapidity
distributions for the largest $y$ values of binned rapidity for the three
analyses are shown in Table~\ref{table:Uncert}.
The contributions include the uncertainty in jet energy scale, unfolding of rapidity distributions, and the scaling of simulated pileup interactions (corresponding to
a 5\% uncertainty in the total inelastic cross section).
For \cPZ\ + jet production, the contributions to the relative experimental
uncertainties also include the uncertainty in lepton identification
efficiency, dominated by the limited statistical precision of simulated event
samples, and the uncertainty in the background subtraction.

For photon production, bounds on the systematic uncertainty in
the modeling of background are determined from the difference
between data and the \PYTHIA simulation in two-jet events.
The maximum extent of the \ydif measurement for photon
production is limited by the number of simulated events used to estimate the
uncertainty from background. The systematic uncertainty in the photon
background as a function of \ydif varies between 2\% and 11\%, and is highly
correlated across the range of \ydif. There is also an uncorrelated
uncertainty from the statistical precision in the estimated background that is
included in the ``Statistical" row of Table~\ref{table:Uncert}.

The best linear unbiased estimator (BLUE)~\cite{Lyons:1998,Valassi:2003mu}
method is used to combine the corrected distributions for the electron and muon decay channels.
A covariance matrix of $2N\times 2N$ dimensions, where 2 refers to the
measurements for the electron and the muon channels and $N$ is the number of
bins, is ascribed to each distribution. The diagonal variances represent the
quadratic sum of both correlated and uncorrelated uncertainties. The off-diagonal
elements are defined by 100\% correlated uncertainties associated
with the jet energy scale in the electron and muon channels.
The size of the uncorrelated uncertainty in each bin is a factor of 3--5 times
greater than the size of the correlated uncertainty.

The sources of theoretical uncertainty in all three analyses include
the choice of PDF, the value of the strong coupling ($\alpha_{\mathrm{s}}$),
and \mur and \muf. These are studied using the \MCFM and Owens
calculations.
The uncertainty in the PDF is obtained using the
PDF4LHC~\cite{PDF4LHC:2011} prescription.
The baseline set of PDF is CTEQ6.6M, while the CT10~\cite{CT10},
MSTW2008~\cite{MSTW}, and NNPDF21~\cite{NNPDF21} PDF are used as alternatives.
For both \cPZ\ + jet and $\gamma$ + jet events, these alternatives
correspond to ${<}2\%$ shift in all $y$ distributions;
however, using the LO rather than NLO parameterizations of PDF produces
${\approx}10\%$ difference in the \ysum and $\abs{\yv}$ distributions, whereas the
\ydif and $\abs{\yj}$ distributions are affected only slightly.
Changes of $\alpha_{\mathrm{s}}$ within the CT10 bounds cause ${<}1\%$
change in all variables.
The theoretical uncertainty from the choice of \muf and \mur is estimated by
changing these scales up and down by a factor of two.
For \cPZ\ + jet and $\gamma$ + jet events, the differences in the
normalized distributions as a function of \ysum, $\abs{\yj}$, and
$\abs{\yv}$ are ${<}2\%$, while the change in \ydif is ${\approx} 8\%$.
Moreover, the changes are similar for both LO and NLO calculations, although
the normalization factor can be quite different.
The NLO calculations do not include effects from final-state photon
radiation, parton showering, and hadronization, since these are estimated
as negligible using the \PYTHIA program.

\begin{figure*}
\centering
\includegraphics[width=\cmsFigWidth]{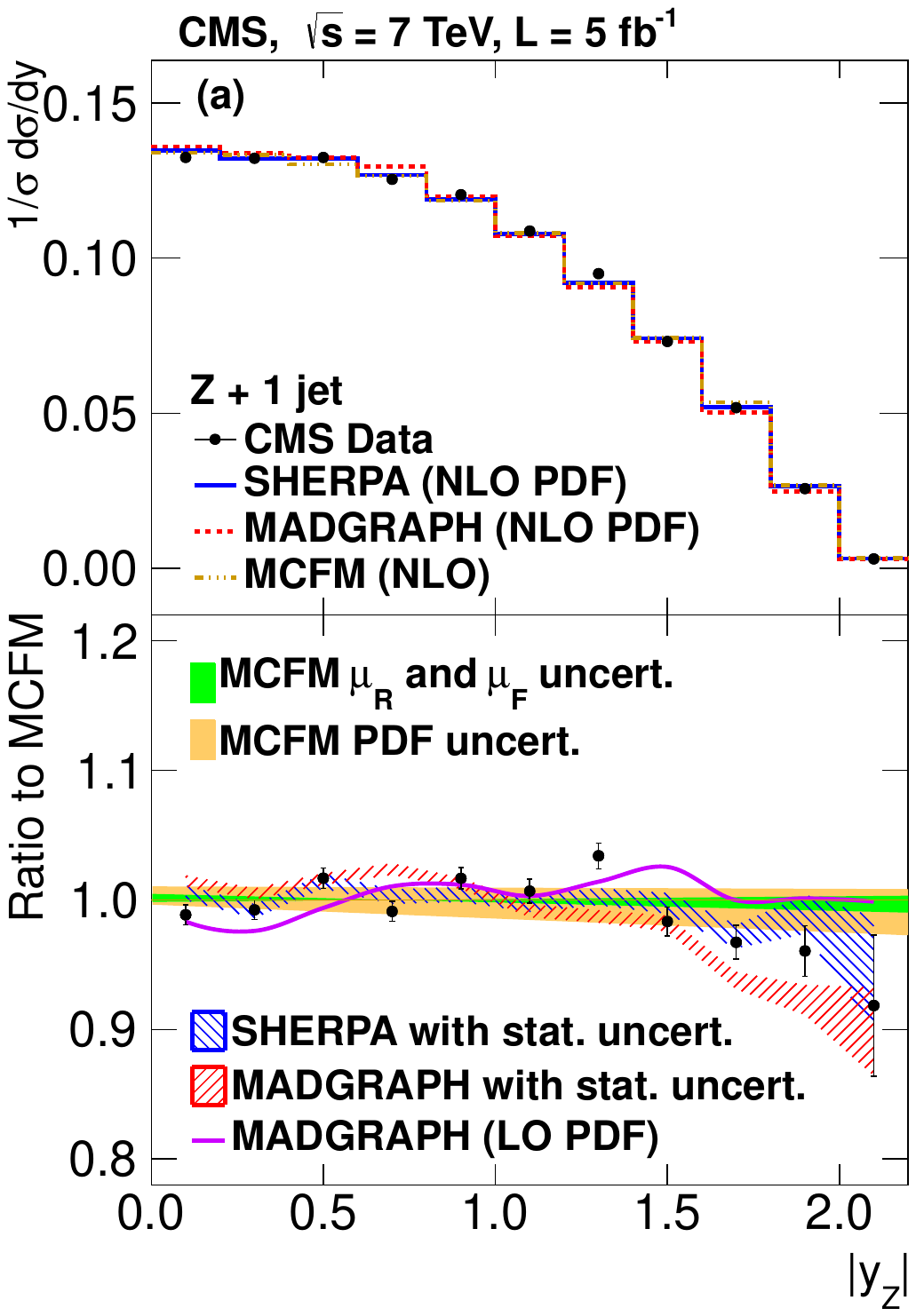}
\includegraphics[width=\cmsFigWidth]{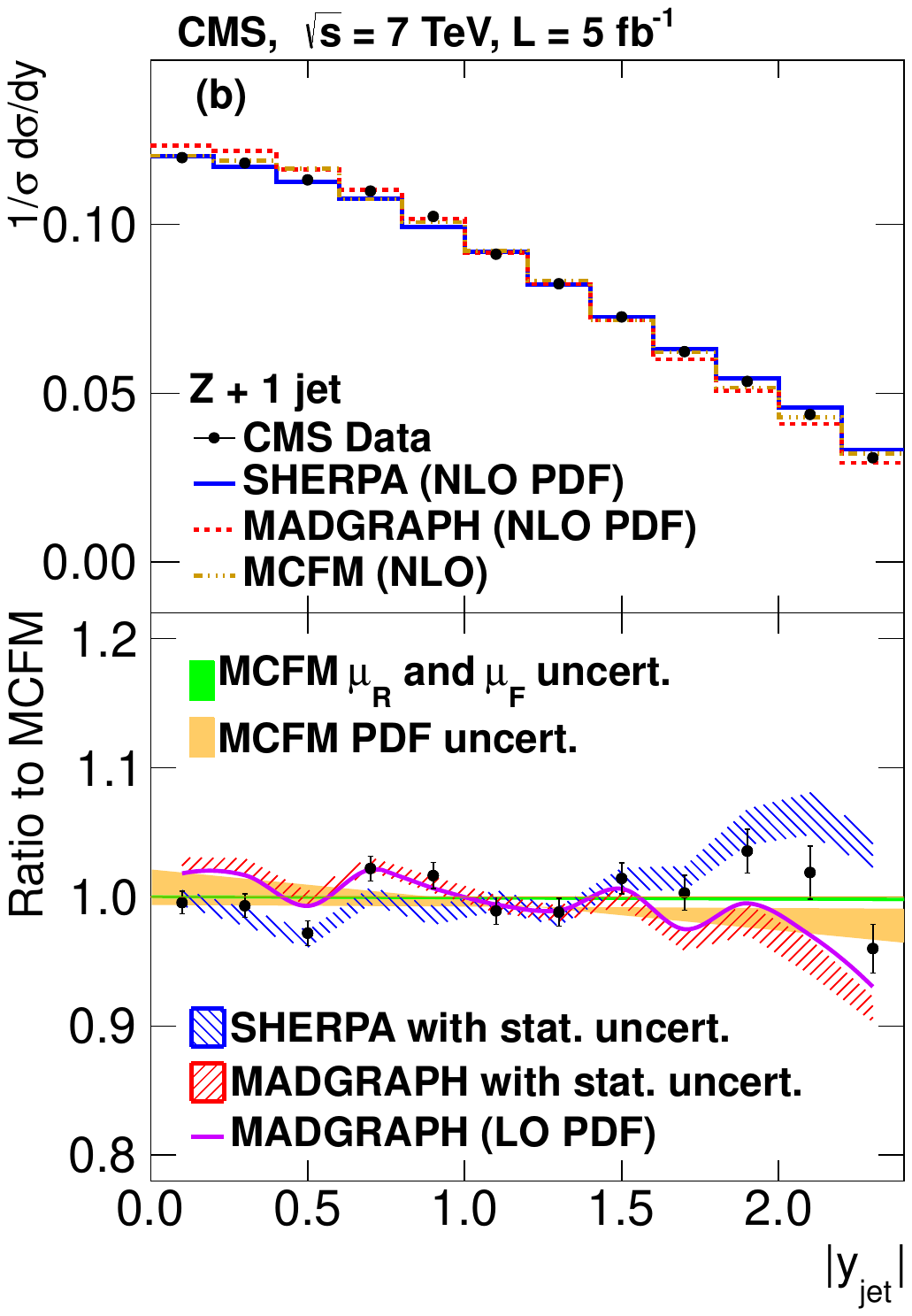} 
\includegraphics[width=\cmsFigWidth]{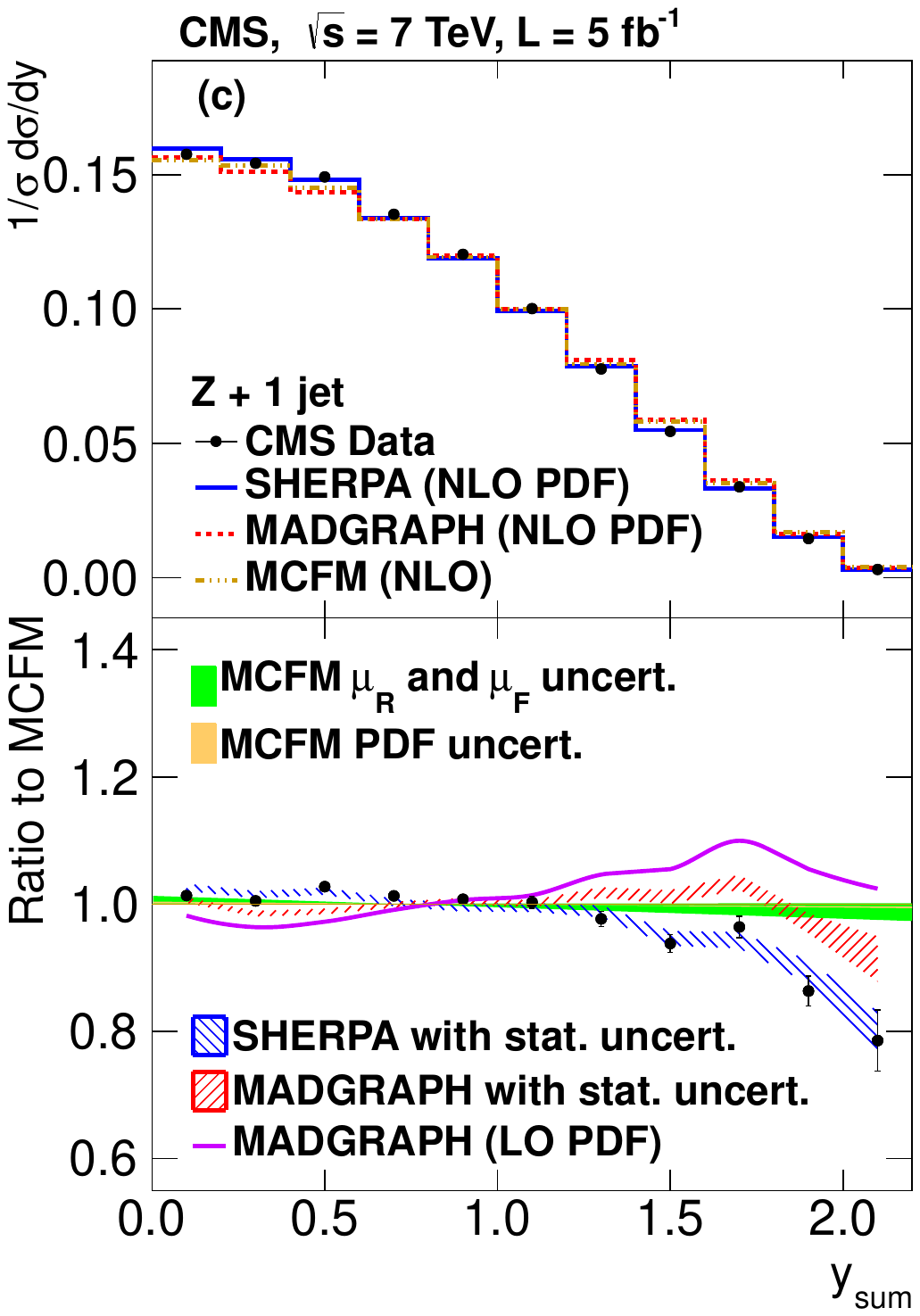}
\includegraphics[width=\cmsFigWidth]{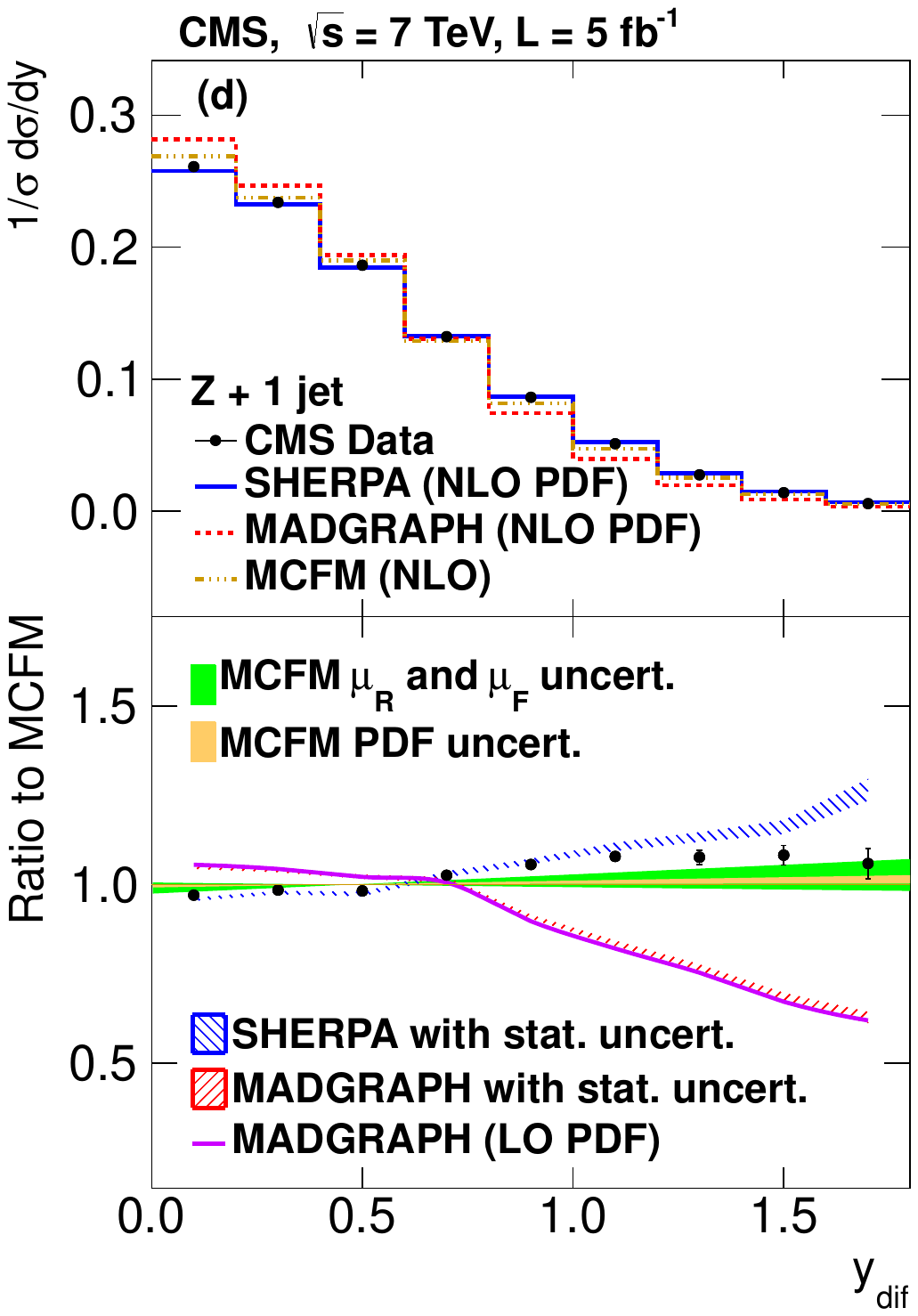}
\caption{Distributions in absolute values of rapidities for
(a)~the \cPZ\ boson, (b)~the jet, (c)~their sums, and (d)~their differences,
normalized to unity. The data are shown after correcting for efficiency and
resolution, and displayed with statistical and systematic uncertainties
combined in quadrature.
The lower panel of each figure gives ratios of the
data and simulations to the NLO calculation of MCFM.
The ratio error bars include MCFM statistical uncertainties folded with data statistical and systematic uncertainties.
Theoretical uncertainties in the MCFM calculations are shown as shaded areas representing variations of \mur, \muf,
and PDF. Statistical uncertainties for the \MADGRAPH and \SHERPA predictions
are displayed as bands around the central values. The central value for
\MADGRAPH simulations using LO PDF is depicted by a line.
All other calculations use NLO versions of PDF.}
 \label{fig:zjet}
\end{figure*}

\begin{figure*}
\centering
\includegraphics[width=\cmsFigWidth]{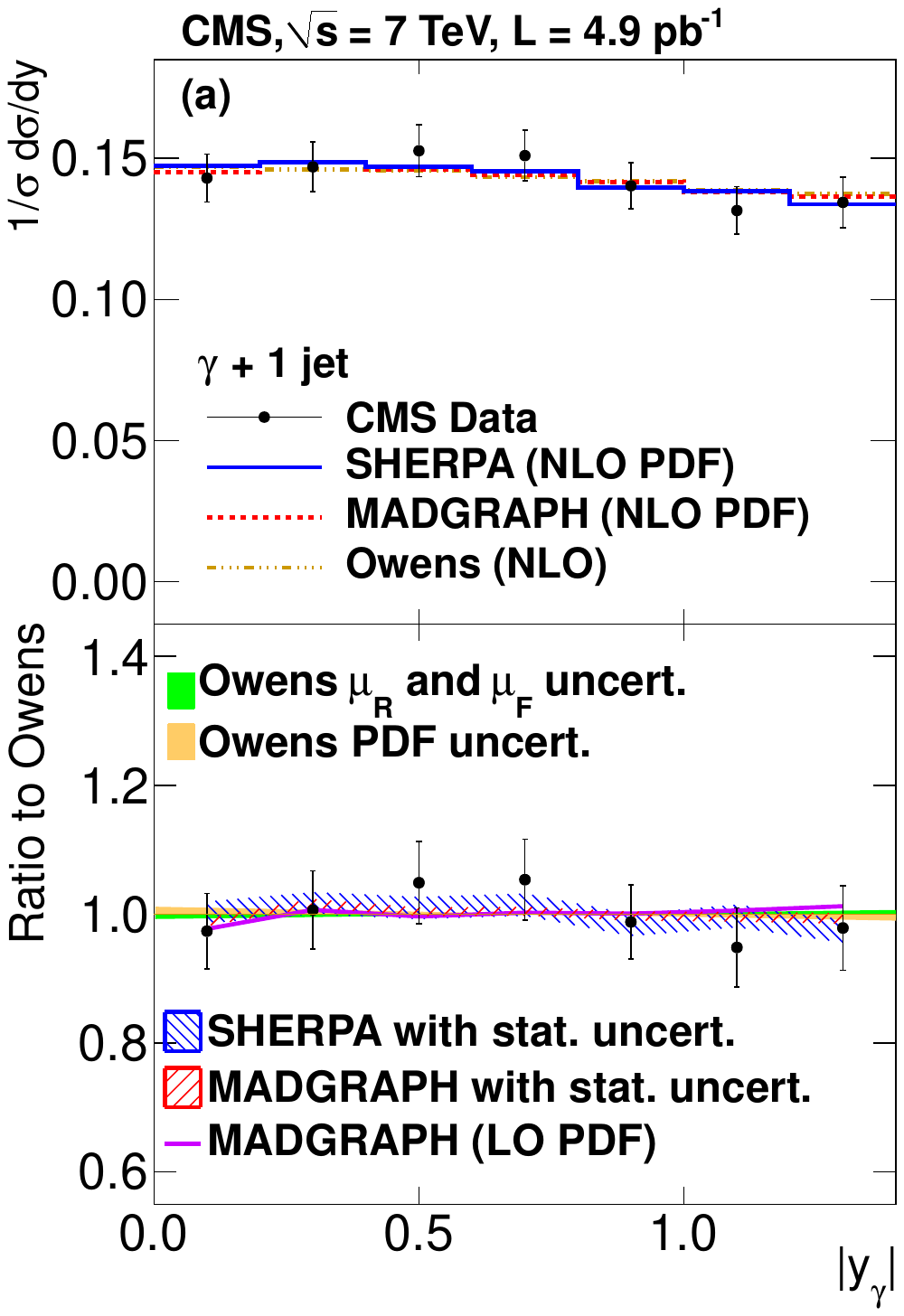} 
\includegraphics[width=\cmsFigWidth]{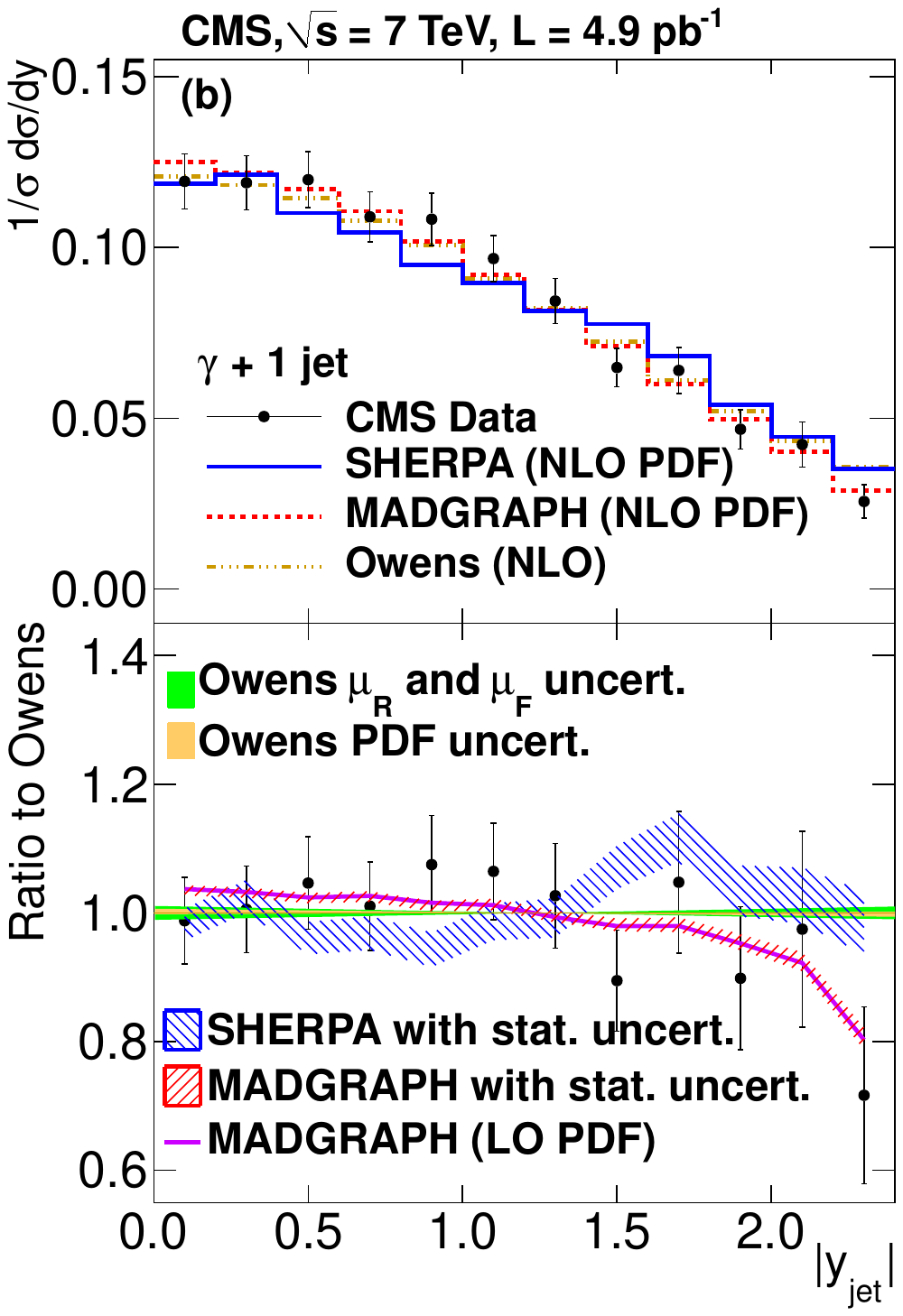}
\includegraphics[width=\cmsFigWidth]{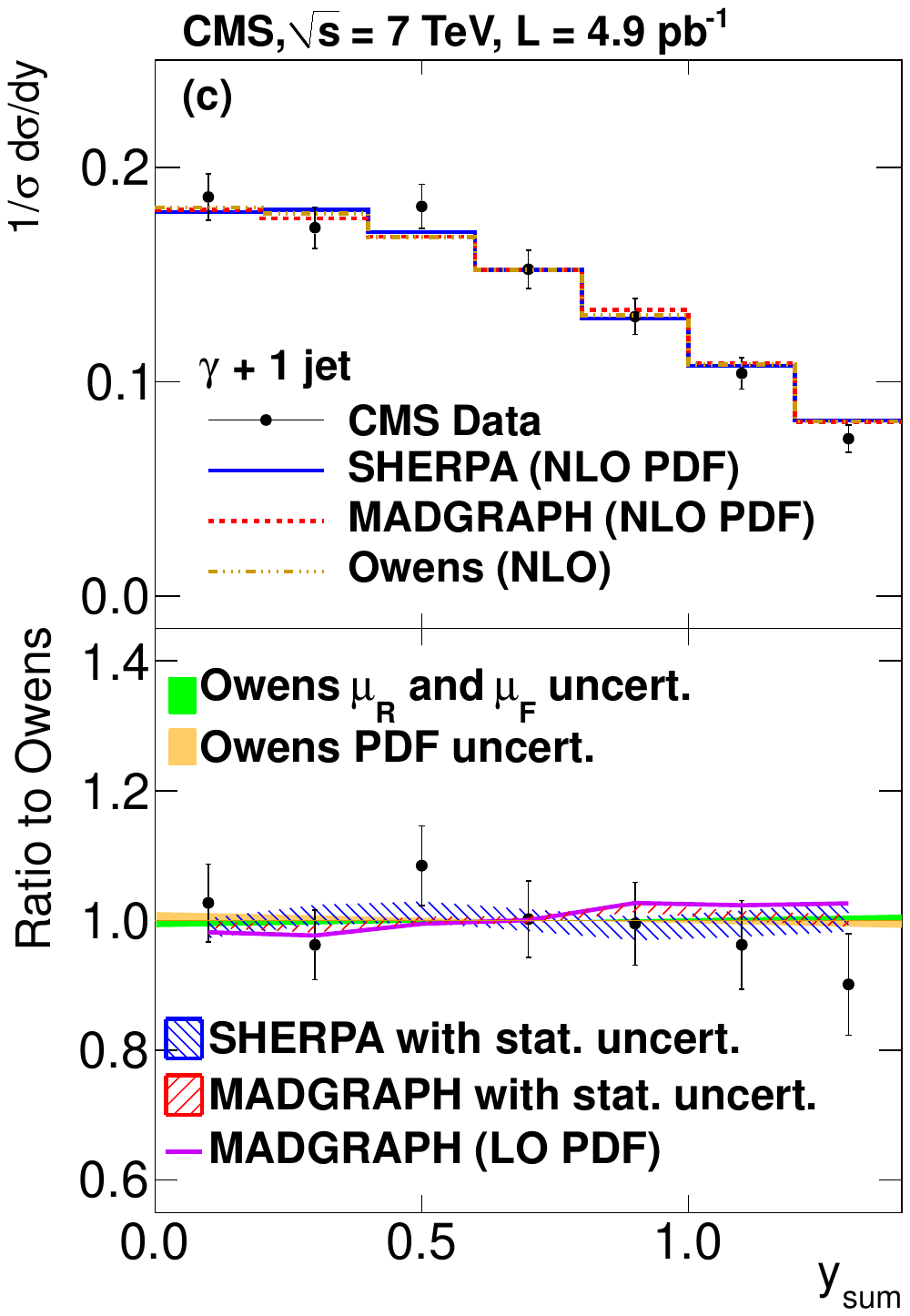}
\includegraphics[width=\cmsFigWidth]{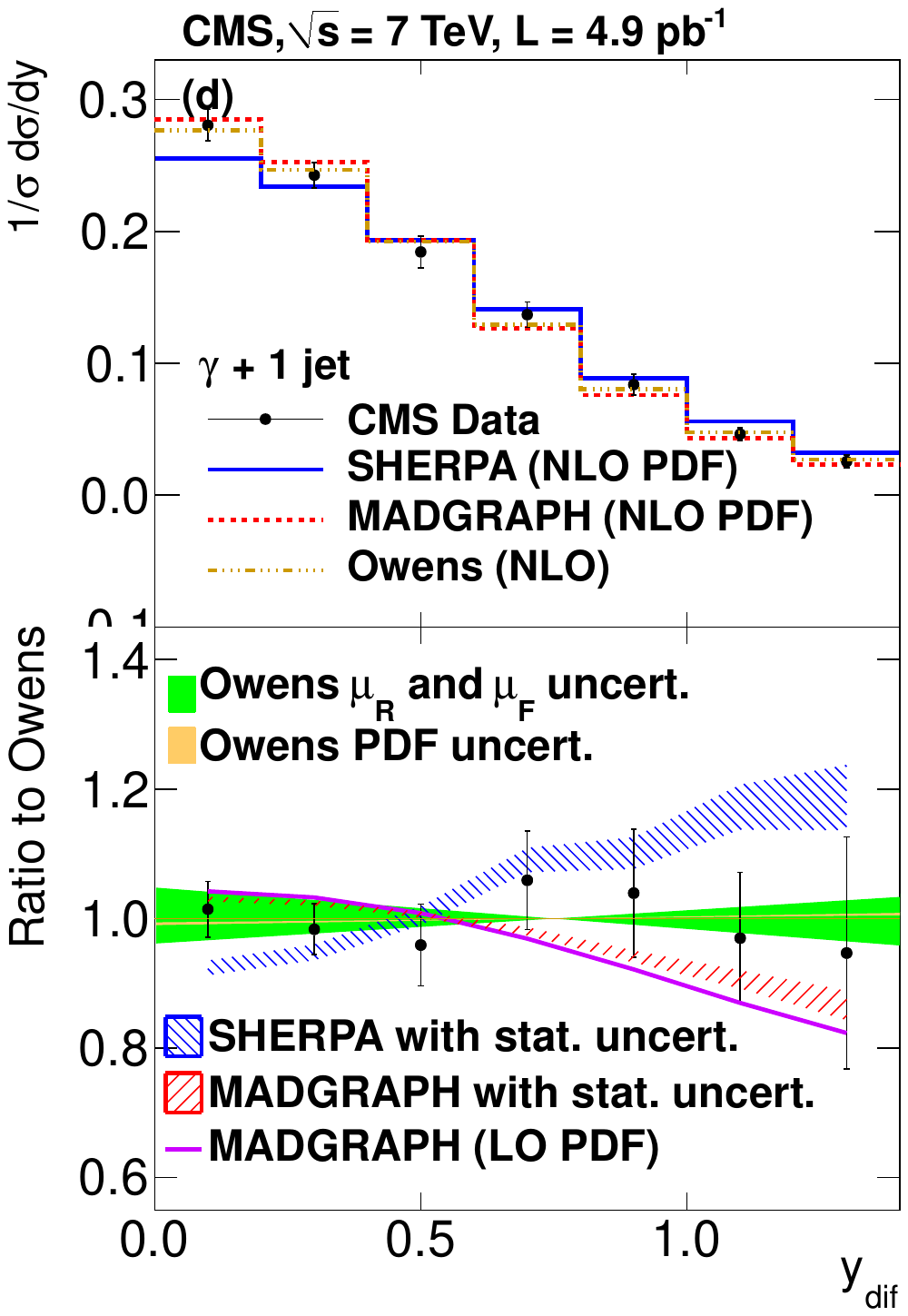}
\caption{Distributions in absolute values of rapidities for
(a)~the photon, (b)~the jet, (c)~their sums, and (d)~their differences,
normalized to unity. The data are shown after correcting for efficiency
and resolution, and displayed with statistical and systematic uncertainties
combined in quadrature. The lower panel of each figure gives ratios of the
data and simulations to the NLO calculation of Owens.
The ratio error bars include Owens statistical uncertainties folded with data statistical and systematic uncertainties.
Theoretical uncertainties in the Owens calculations are shown as
shaded areas representing variations of \mur, \muf,  and PDF. Statistical
uncertainties for the \MADGRAPH and \SHERPA predictions are displayed as
 bands around the central values. The central value for \MADGRAPH simulations
using LO PDF is depicted by a line.
All other calculations use NLO versions of PDF.}
\label{fig:gjet}
\end{figure*}

The normalized rapidity distributions for $\abs{\yz}$, $\abs{\yj}$, \ydif, and \ysum,
along with predictions from theory, are shown in Fig.~\ref{fig:zjet} for \cPZ\ +
jet events. The data for the $\abs{\yz}$ and $\abs{\yj}$ distributions agree to better
than 5\% accuracy with \SHERPA, \MADGRAPH, and \MCFM over the full range of
the measurement.
The \SHERPA simulation reproduces the features of the \ysum distribution
better than the \MADGRAPH or \MCFM programs.
As shown in Fig.~\ref{fig:zjet},  when \MADGRAPH events are simulated using
the LO PDF, the distributions of \ysum and $|\yz|$ are less consistent with
the data.
The \ydif distribution is consistent with  \MCFM for $\ydif < 1.0$.
As was noted in Ref.~\cite{Alwall:2008}, the two hybrid programs differ
considerably in the prediction for \ydif.
Since both \MADGRAPH and \SHERPA use the same LO matrix elements and
approaches to parton showering, the difference in the distribution of \ydif
can be attributed to the matching algorithm, with the \SHERPA CKKW scheme
appearing more consistent with the data.
Indeed, the \MADGRAPH distribution of \ydif resembles the LO distribution.
The difference in \ydif between the LO and NLO calculations is due to the
contribution from NLO diagrams with a gluon propagator that yield more
forward rapidities. The rapidity distributions for $\gamma$ + jet events
shown in Fig.~\ref{fig:gjet} are consistent with perturbative QCD.
The qualitative difference in \ydif for the hybrid generators
\MADGRAPH and \SHERPA is comparable to that observed in \cPZ\ + jet events,
although the statistical precision of the $\gamma$ + jet measurement is
insufficient to discriminate between the theoretical alternatives.

In summary, the CMS detector was used to measure the rapidities of
particles in events containing a vector (V) boson (either a \cPZ\ boson or photon)
in association with a single jet in $\Pp\Pp$ collisions at $\sqrt{s} = 7\TeV$
for an integrated luminosity of $5.0\fbinv$. The rapidity distributions of
$\abs{\yv}$ and $\abs{\yj}$
are found to agree with predictions from the  \SHERPA, \MADGRAPH, and \MCFM
QCD models.
The distribution for the sum of the \Vvar  and jet rapidities is described by all
predictions to better than 5\% precision for $\ysum<1.0$, and is best
described by hybrid calculations that employ NLO PDF. The distribution in the
difference in rapidities (\ydif) is described to better than 10\% by the
\MCFM prediction.
The two hybrid event generators differ by as much as ${\approx}50\%$ in their
predictions for distributions in \ydif, but \SHERPA is significantly closer to
data than \MADGRAPH. Nevertheless, \SHERPA overestimates the cross section at
higher \ydif. We attribute the difference between the hybrid event generator distributions
to the respective methods by which partons from matrix elements
are matched to parton showers.

The authors thank  J.~Campbell, R.~Frederix, and J.~Owens for their
substantial contributions to this work.
We congratulate our colleagues in the CERN accelerator departments for
the excellent performance of the LHC and thank the technical and
administrative staffs at CERN and at other CMS institutes for their
contributions to the success of the CMS effort. In addition, we
gratefully acknowledge the computing centers and personnel of the
Worldwide LHC Computing Grid for delivering so effectively the
computing infrastructure essential to our analyses. Finally, we
acknowledge the enduring support for the construction and operation of
the LHC and the CMS detector provided by the following funding
agencies: BMWF and FWF (Austria); FNRS and FWO (Belgium); CNPq, CAPES,
FAPERJ, and FAPESP (Brazil); MEYS (Bulgaria); CERN; CAS, MoST, and
NSFC (China); COLCIENCIAS (Colombia); MSES (Croatia); RPF (Cyprus);
MoER, SF0690030s09 and ERDF (Estonia); Academy of Finland, MEC, and
HIP (Finland); CEA and CNRS/IN2P3 (France); BMBF, DFG, and HGF
(Germany); GSRT (Greece); OTKA and NKTH (Hungary); DAE and DST
(India); IPM (Iran); SFI (Ireland); INFN (Italy); NRF and WCU
(Republic of Korea); LAS (Lithuania); CINVESTAV, CONACYT, SEP, and
UASLP-FAI (Mexico); MSI (New Zealand); PAEC (Pakistan); MSHE and NSC
(Poland); FCT (Portugal); JINR (Armenia, Belarus, Georgia, Ukraine,
Uzbekistan); MON, RosAtom, RAS and RFBR (Russia); MSTD (Serbia); SEIDI
and CPAN (Spain); Swiss Funding Agencies (Switzerland); NSC (Taipei);
ThEPCenter, IPST and NSTDA (Thailand); TUBITAK and TAEK (Turkey); NASU
(Ukraine); STFC (United Kingdom); DOE and NSF (USA).

\bibliography{auto_generated}

\cleardoublepage \appendix\section{The CMS Collaboration \label{app:collab}}\begin{sloppypar}\hyphenpenalty=5000\widowpenalty=500\clubpenalty=5000\textbf{Yerevan Physics Institute,  Yerevan,  Armenia}\\*[0pt]
S.~Chatrchyan, V.~Khachatryan, A.M.~Sirunyan, A.~Tumasyan
\vskip\cmsinstskip
\textbf{Institut f\"{u}r Hochenergiephysik der OeAW,  Wien,  Austria}\\*[0pt]
W.~Adam, T.~Bergauer, M.~Dragicevic, J.~Er\"{o}, C.~Fabjan\cmsAuthorMark{1}, M.~Friedl, R.~Fr\"{u}hwirth\cmsAuthorMark{1}, V.M.~Ghete, N.~H\"{o}rmann, J.~Hrubec, M.~Jeitler\cmsAuthorMark{1}, W.~Kiesenhofer, V.~Kn\"{u}nz, M.~Krammer\cmsAuthorMark{1}, I.~Kr\"{a}tschmer, D.~Liko, I.~Mikulec, D.~Rabady\cmsAuthorMark{2}, B.~Rahbaran, C.~Rohringer, H.~Rohringer, R.~Sch\"{o}fbeck, J.~Strauss, A.~Taurok, W.~Treberer-Treberspurg, W.~Waltenberger, C.-E.~Wulz\cmsAuthorMark{1}
% ...
\vskip\cmsinstskip
% ...
1:~~Also at Vienna University of Technology, Vienna, Austria\\
2:~~Also at CERN, European Organization for Nuclear Research, Geneva, Switzerland\\
% ...

\end{sloppypar}
\end{document}